\documentclass[prl,reprint]{revtex4-1}
\pdfoutput=1
\usepackage{graphicx}
\usepackage{color}
\usepackage{bm}
\usepackage{hyperref}

\begin{document}

\title{Observation of an Anisotropic Wigner Crystal}
\date{today}

\author{Yang Liu}
\affiliation{Department of Electrical Engineering,
Princeton University, Princeton, New Jersey 08544}
\author{S.\ Hasdemir}
\affiliation{Department of Electrical Engineering,
Princeton University, Princeton, New Jersey 08544}
\author{L.N.\ Pfeiffer}
\affiliation{Department of Electrical Engineering,
Princeton University, Princeton, New Jersey 08544}
\author{K.W.\ West}
\affiliation{Department of Electrical Engineering,
Princeton University, Princeton, New Jersey 08544}
\author{K.W.\ Baldwin}
\affiliation{Department of Electrical Engineering,
Princeton University, Princeton, New Jersey 08544}
\author{M.\ Shayegan}
\affiliation{Department of Electrical Engineering,
Princeton University, Princeton, New Jersey 08544}

\date{\today}

\begin{abstract}
  We report a new correlated phase of two-dimensional charged carriers
  in high magnetic fields, manifested by an anisotropic insulating
  behavior at low temperatures. It appears near Landau level filling
  factor $\nu=1/2$ in hole systems confined to wide GaAs quantum wells
  when the sample is tilted in magnetic field to an intermediate
  angle. The parallel field component ($B_{||}$) leads to a crossing
  of the lowest two Landau levels, and an elongated hole wavefunction
  in the direction of $B_{||}$. Under these conditions, the in-plane
  resistance exhibits an insulating behavior, with the resistance
  along $B_{||}$ more than 10 times smaller than the resistance
  perpendicular to $B_{||}$. We interpret this anisotropic insulating
  phase as a two-component, striped Wigner crystal.
\end{abstract}

\maketitle

Low-disorder, two-dimensional (2D) systems of charged carriers, cooled
to low temperatures and subjected to a strong perpendicular magnetic
field ($B_{\perp}$) are host to a plethora of exotic, quantum
many-body states \cite{Tsui.PRL.1982, Shayegan.Flatland.2006,
  Jain.CF.2007}. At odd-denominator fractional fillings of the lowest
Landau level (LL), they exhibit fractional quantum Hall states
(FQHSs), uniform-density, incompressible liquid phases for which the
resistance vanishes as temperature $T$ approaches absolute zero
\cite{Tsui.PRL.1982, Shayegan.Flatland.2006, Jain.CF.2007}. On the
other hand, when the filling factor becomes very small
($\nu \lesssim 1/5$), the system condenses into an ordered array of
electrons, the so-called Wigner crystal, which is insulating because
it is pinned by the ubiquitous disorder potential
\cite{Shayegan.Flatland.2006, Jain.CF.2007, Andrei.PRL.1988,
  Jiang.PRL.1990, Goldman.PRL.1990, Shayegan.PQHE.1998,
  Liu.cond.mat.2014}.  Yet another set of states are the anisotropic
phases observed at large even-denominator fillings (e.g., $\nu=9/2$)
which are believed to be nematic liquid states \cite{Lilly.PRL.1999,
  Du.SSC.1999, Shayegan.PhysicaE.2000, Fradkin.ARCMP.2010}. The new
correlated phase we report here is distinct from these states as it
shows an \textit{anisotropic insulating} behavior. It is manifest at
low fillings (near $\nu=1/2$) in 2D hole systems (2DHSs) with a
bilayer charge distribution and tilted in magnetic field to introduce
a field component ($B_{||}$) parallel to the 2D plane. Curiously, the
anisotropic phase forms in a relatively narrow range of tilt angles
near $\theta\simeq35^{\circ}$ when the two lowest energy LLs are very
close in energy. Outside this range, the 2DHS is not insulating and
exhibits FQHSs at numerous fillings. The conditions under which the
new insulating phase appears suggest that it is an anisotropic
(striped), two-component, pinned Wigner crystal (Fig. 1(a)).

\begin{figure*}
\includegraphics[width=.95\textwidth]{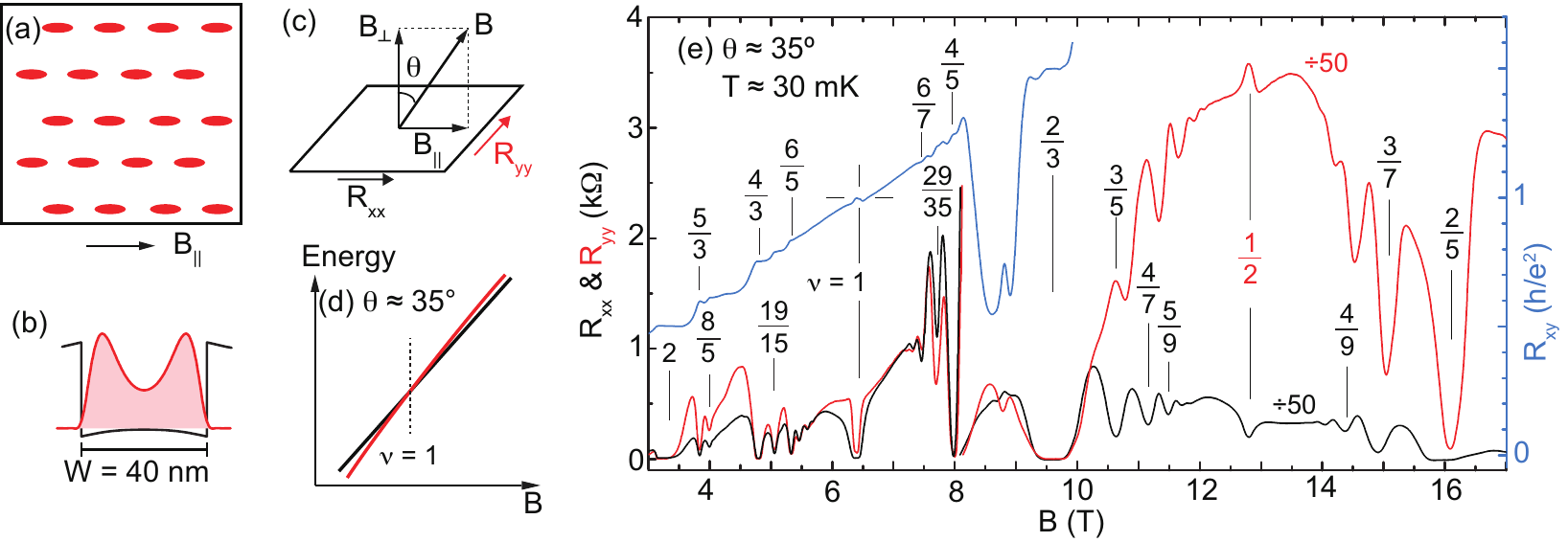}
\caption{(color online) (a) Conceptual plot of an anisotropic
  (striped) Wigner crystal. (b) The charge distribution (red)
  and potential (black), from calculating the Schroedinger and
  Poisson's equations self-consistently at $B=0$. (c) Experimental
  geometry: $R_{xx}$ and $R_{yy}$ denote the longitudinal
  magneto-resistance measured along and perpendicular to the parallel
  magnetic field ($B_{||}$), respectively. (d) Schematic diagram,
  showing the crossing of the lowest two Landau levels at
  $\theta\simeq 35^{\circ}$ near $\nu=1$. (e) Magneto-resistance
  traces measured at tilt angle $\theta\simeq 35^{\circ}$ for a 2DHS
  with density $p=1.28\times 10^{11}$ cm${^{-2}}$ and confined to a
  40-nm-wide GaAs QW. Note the factor of 50 change in the scale for
  $R_{xx}$ and $R_{yy}$ for $B > 8$ T.}
\end{figure*}

Our 2DHSs are confined to 40- and 50-nm-wide GaAs quantum wells (QWs)
flanked by undoped Al$_{0.3}$Ga$_{0.7}$As spacer and C
$\delta$-doped layers, and have as grown densities
$\simeq 1.2\times 10^{11}$ cm$^{-2}$. The structures were grown by
molecular beam epitaxy on GaAs (001) wafers and have very high
low-temperature mobilities, $\mu \gtrsim$ 100 m$^2$/Vs. Each sample has a
4 $\times$ 4 mm$^2$ van der Pauw geometry with alloyed In:Zn contacts
at its four corners. We then fit it with an evaporated Ti/Au
front-gate and an In back-gate to control the 2DHS density ($p$) and
keep the QW symmetric. The holes in the QW have a bilayer-like charge
distribution (Fig. 1(b)). The transport measurements were carried out
in a dilution refrigerator with a base temperature of $T \simeq$ 30
mK, and an \textit{in-situ} rotatable sample platform to induce
$B_{||}$. As illustrated in Fig. 1(c), we use $\theta$ to express the
angle between the field and the sample plane normal, and denote the
longitudinal resistances measured along and perpendicular to the
direction of $B_{||}$ by $R_{xx}$ and $R_{yy}$, respectively. We used
low-frequency ($\sim 30$ Hz) lock-in technique to measure the
transport coefficients.

Figure 1 (e) highlights our main finding. It shows the longitudinal
($R_{xx}$ and $R_{yy}$) and Hall ($R_{xy}$) magneto-resistance traces,
measured for a 2DHS confined to a 40-nm-wide GaAs QW at
$p=1.28\times 10^{11}$ cm$^{-2}$ and $\theta\simeq 35^{\circ}$.
Starting at $B\simeq 8$ T, both $R_{xx}$ and $R_{yy}$ rapidly
increase; note the 50 times change of scale for $R_{xx}$ and $R_{yy}$
above 8 T \cite{Note1, Goldman.PRL.1993, Sajoto.PRL.1993}. 
Most remarkably, near $\nu=1/2$, $R_{xx}$ is $\sim$ 25
k$\Omega$ while $R_{yy}\simeq 10 R_{xx}$ and,
as we will show shortly, both $R_{xx}$ and $R_{yy}$ exhibit an
insulating behavior. To probe the origin of this anisotropic
insulating phase (IP), we present several experimental observations.

Data of Fig. 2, which were taken at $\theta=0$ and $50^{\circ}$,
demonstrate that the anisotropic IP seen in Fig. 1(e) occurs near a
crossing of the lowest two LLs (Fig. 1(d)). The crossing is
signaled by a profound weakening of the $\nu=1$
integer QHS at intermediate $\theta$ \cite{Graninger.PRL.2011,
  Liu.PRB.2015B}. As is evident in Fig. 2, traces taken at both
$\theta=0$ and $50^{\circ}$ show a strong integer QHS at $\nu=1$, with a
very wide resistance plateau and large excitation gap
$\Delta\gtrsim 10$ K. In contrast, at intermediate angle
$\theta\simeq35^{\circ}$ (Fig. 1(e)), the $\nu=1$ QHS becomes much
weaker ($\Delta\simeq 0.22$ K) and has a very narrow plateau.

The evolution of the FQHSs in Figs. 1(e) and 2 are also consistent
with a LL crossing occurring near $\nu=1$ when
$\theta\simeq35^{\circ}$. In Fig. 2 traces there are numerous strong
FQHSs at the well-known, ``standard'' $\nu=i/(2i\pm 1)$ fillings
($i>0$ is an integer) \cite{Jain.CF.2007}. In Fig. 1(e) data, however,
near $\nu=1$ there are uncharacteristically strong FQHSs at the
\textit{even-numerator} fillings $\nu=4/3$, 6/5, 6/7, and 4/5. This is
similar to what is seen in bilayer 2D \textit{electron} systems
(2DESs) with extremely small energy separation between the lowest two
LLs \cite{Manoharan.PRL.1997}, and implies that these are
two-component FQHSs, each component having half of the total
filling. In Fig. 1(e) we also observe FQHSs at very unusual fillings
such as $\nu=19/15$ and 29/35. Such states were seen in
Ref. \cite{Manoharan.PRL.1997} when the lowest two LLs are nearly
degenerate, and were interpreted as ``imbalanced'' two-component
FQHSs: for example, the $\nu=19/15$ FQHS has fillings 2/3 and 3/5 for
its two components. In Fig. 2, we also observe strong FQHSs at the
even-denominator filling $\nu=1/2$. This FQHS is seen in 2DESs and
2DHSs confined to wide GaAs QWs \cite{Suen.PRL.1992, Suen.PRL.1994,
  Shabani.PRB.2013, Liu.PRL.2014}. It is likely the $\Psi_{331}$
state, a two-component FQHS stabilized by strong and comparable
interlayer and intralayer interactions which are prevalent at
$\nu=1/2$ \cite{Note2}.

\begin{figure}
\includegraphics[width=.45\textwidth]{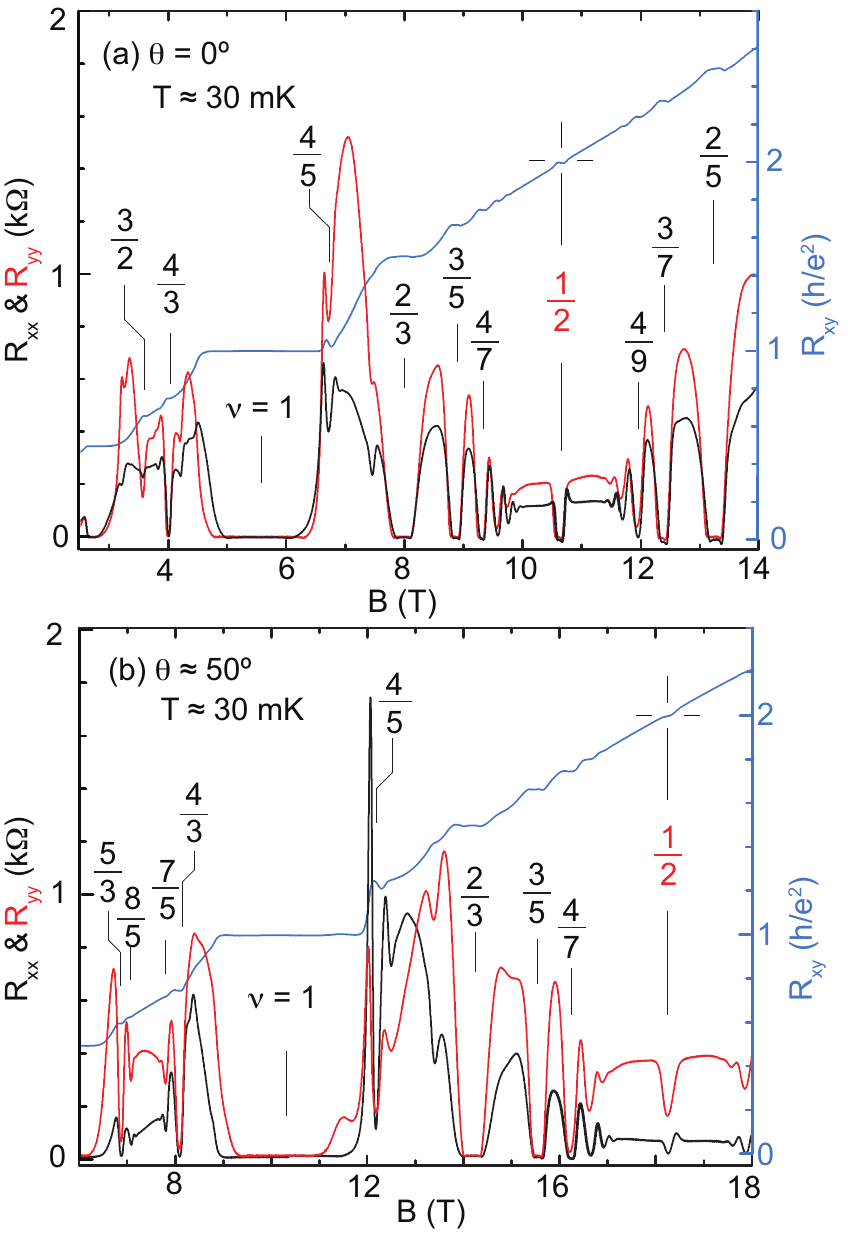}
\caption{(color online) Magneto-resistance traces for the sample of
  Fig. 1, measured at $\theta=0^{\circ}$ and 50$^{\circ}$. Both
  $R_{xx}$ and $R_{yy}$ are much smaller near $\nu=1/2$ than in
  Fig. 1(e) data.}
\end{figure}

\begin{figure}
\includegraphics[width=.45\textwidth]{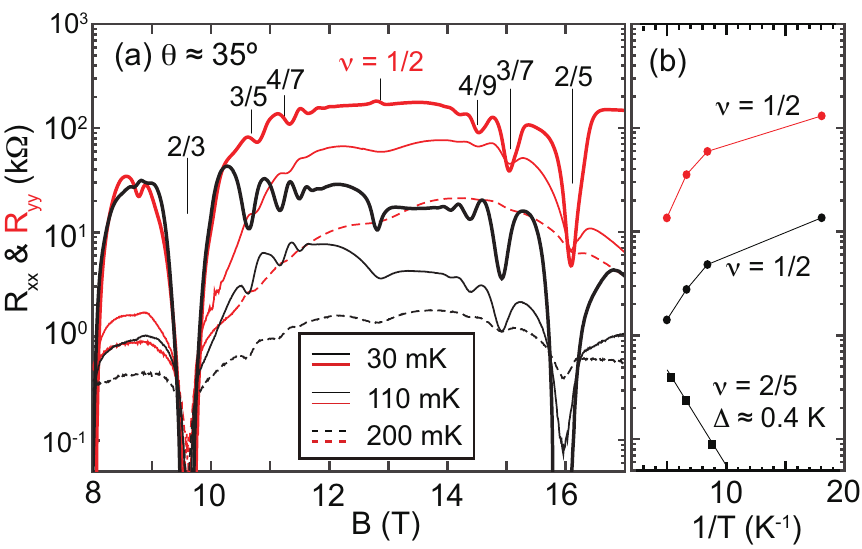}
\caption{(color online) (a) ${R_{xx}}$ and $R_{yy}$ measured at three
  temperatures in the 40-nm-wide QW near $\nu=1/2$ at
  $\theta\simeq 35^{\circ}$. (b) $T$-dependence of $R_{xx}$ and
  $R_{yy}$ at $\nu=1/2$. To avoid the influence of the relatively
  sharp $\nu=1/2$ resistance minima seen in some of the traces, here
  we are plotting the “background” resistances by interpolating
  between the shoulders flanking the $\nu=1/2$ minima. Also plotted is
  the $T$-dependence of the $R_{xx}$ minimum for the $\nu=2/5$ FQHS.}
\end{figure}

Next we focus on the anisotropic IP seen at $\theta\simeq 35^{\circ}$.
Figure 3 captures the insulating behavior of this phase near
$\nu=1/2$. Both $R_{xx}$ and $R_{yy}$ increase as temperature is
decreased but, as can be best seen in Fig. 3(b), $R_{yy}$ is about 10
times larger than $R_{xx}$. Before discussing this anisotropic
behavior, it is instructive to first briefly review the IPs seen in 2D
systems at low fillings.

It is well established that in very clean 2D systems of charged
carriers, at very small $\nu$ ($\nu\lesssim 1/5$ for electrons and
$\nu\lesssim 1/3$ for holes) the FQHSs give way to IPs which are
believed to be Wigner crystal (WC) phases that are pinned by the small
but ubiquitous disorder \cite{Lozovik.JETP.1975, Andrei.PRL.1988,
  Jiang.PRL.1990, Goldman.PRL.1990, Santos.PRL.1992, Santos.PRB.1992,
  Li.PRL.1997, Shayegan.PQHE.1998}. For both 2D electron and hole
systems in wide, symmetric GaAs QWs, the charge distribution becomes
more bilayer-like with increasing density and the IP sets in at
progressively larger $\nu$ \cite{Manoharan.PRL.1996, Liu.PRL.2014,
  Hatke.Nat.Comm.2015}. These IPs are believed to be \textit{bilayer}
WC states which, thanks to the additional layer/subband degree of
freedom, are stabilized at relatively large $\nu$ compared to the
single-layer systems. For example, in 2DHSs confined to a 40-nm-wide
QW with $p\gtrsim 1.7\times 10^{11}$ cm$^{-2}$ \cite{Liu.PRL.2014}, an
IP is observed near $\nu=1/2$. All the IPs described above are
\textit{isotropic}, and were observed in the absence of $B_{||}$
\cite{Note3, Hasdemir.PRB.2015}.

To discuss the likely origin of the anisotropic IP we observe near
$\nu=1/2$, we focus on its key attributes:

\textit{(i) It is a collective state.}  There are numerous FQHSs near
$\nu=1/2$ in Fig. 1(e), e.g., at $\nu$ = 2/5, 3/7, 4/9, 3/5, 4/7,
5/9. These correlated states are much weaker than the FQHSs seen at
the same fillings in Fig. 2 traces, but their mere presence in
Fig. 1(e) strongly suggests that correlations are prevalent near
$\nu=1/2$ where the IP reigns. Also worth emphasizing is that in
Fig. 2 traces there are very strong FQHSs near and even \textit{at}
$\nu=1/2$. It is very unlikely that at the intermediate tilt angle of
Fig. 1(e) interactions would disappear and the ground state become of
single-particle origin.

\textit{(ii) It is a two-component state.}  It is
  clear that the anisotropic IP is observed near a LL crossing, and
  Fig. 2 traces, which were taken far from the LL crossing, do not
  show insulating behavior near $\nu=1/2$. This implies that the
  presence of two nearly degenerate LLs plays a crucial role for its
  stability. Also, theoretical calculations rule out any
  single-component WC near $\nu=1/2$ \cite{Archer.PRL.2013}. The
  anisotropic IP we observe at $\theta\simeq 35^{\circ}$ is thus
  likely to have a two-component origin. In Fig. 3(a), the existence
  of minima at $\nu = 2/5$, 3/7, 3/5, and their deepening (relative to
  the insulating background resistance) at lower temperatures, signal
  a close competition between a reentrant two-component WC phase and
  the FQHSs. We add that, whether isotropic or anisotropic,
  one-component or two-component, the observation of an IP near the
  crossing of two LLs is by itself unprecedented.

\begin{figure}[htb]
\includegraphics[width=.4\textwidth]{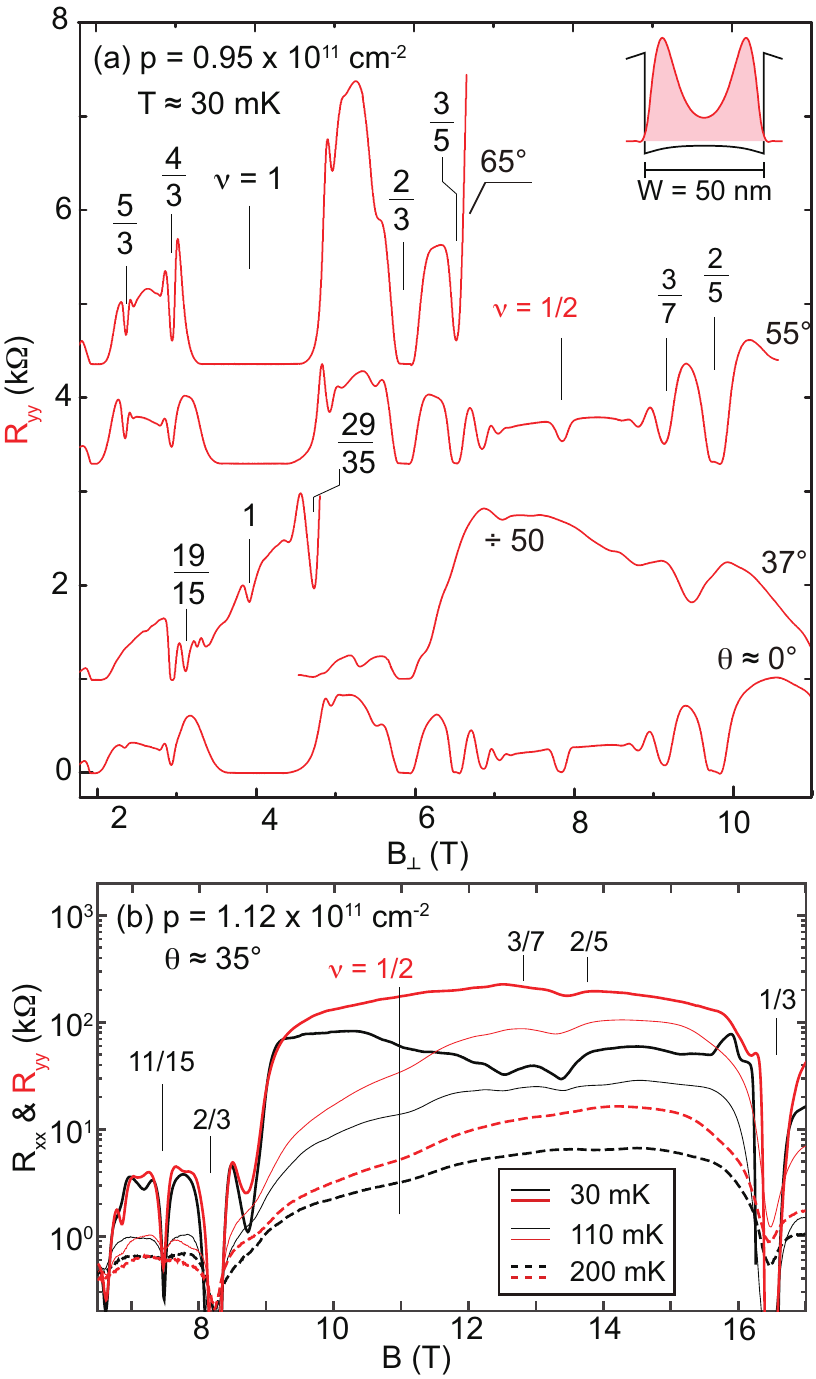}%
\caption{(color online) (a) $R_{yy}$ vs $B_{\perp}$ traces measured for
  a 2DHS confined to a 50-nm-wide GaAs QW at density
  $p=0.95\times 10^{11}$ cm${^{-2}}$ and at different $\theta$. The
  2DHS exhibits an insulating behavior near $\nu=1/2$ at
  $\theta\simeq 37^{\circ}$ and at $\theta=65^{\circ}$, but not at
  $\theta = 25^{\circ}$ or $\theta=55^{\circ}$. The inset shows the
  calculated charge distribution and potential at $B=0$. (b) $R_{xx}$
  and $R_{yy}$ traces are shown at $\theta\simeq 35^{\circ}$ and at a
  slighlty larger density ($p=1.12\times 10^{11}$ cm${^{-2}}$) for
  different temperatures. Similar to the data of Fig. 3, both
  $R_{xx}$ $R_{yy}$ exhibit insulating behavior and the system is
  anisotropic near $\nu=1/2$. }
\end{figure}

\textit{(iii) It is not a nematic liquid state.} One might naively
conclude that the anisotropy we report resembles the one observed at
higher half-filled LLs, and believed to signal nematic electron phases
\cite{Lilly.PRL.1999, Du.SSC.1999, Shayegan.PhysicaE.2000, Fradkin.PRB.1999,
  Fradkin.ARCMP.2010}. But this is incorrect, as there are two major,
qualitative differences. First, nematic phases are \textit{liquid}
states: while in-plane transport becomes anisotropic, $R_{xx}$ and
$R_{yy}$ do not diverge at low temperatures; instead they remain
finite and in fact the resistance along the ``easy axis'' direction
decreases as temperature is lowered and attains extremely small values
\cite{Lilly.PRL.1999, Du.SSC.1999}. This is very different from the
\textit{insulating} behavior and the large values we measure for both
$R_{xx}$ and $R_{yy}$. Second, in the case of $B_{||}$-induced nematic
phases, the ``hard axis'' is typically along $B_{||}$
\cite{Pan.PRL.1999, Lilly.PRL.83.1999}. We observe the opposite
behavior: the resistance along $B_{||}$ ($R_{xx}$) is \textit{smaller} than in
the perpendicular direction \cite{Note4}. 

Based on the above observations, we associate the IP observed in
Fig. 1(e) with a pinned, anisotropic WC. We suggest that the
anisotropy originates from the strongly distorted shape of the hole
charge distribution induced by $B_{||}$, as schematically depicted in
Fig. 1(a). Because of the finite thickness of the hole layer in our
sample, $B_{||}$ couples to the out-of-plane (orbital) motion of the
carriers, and squeezes the charge distribution in the direction
perpendicular to $B_{||}$ (see Fig. 1(a)). Such distortions have been
recently documented for carriers near $B_{\perp}=0$, and also for
composite fermions at high $B_{\perp}$ \cite{Kamburov.PRB.2012,
  Kamburov.PRL.2013, Kamburov.PRB.2014}. An elongated charge
distribution can lead to anisotropic interaction, and provides a
natural explanation for the anisotropic IP we observe in terms of a
pinned, striped WC as shown in Fig. 1(a) \cite{Note5, Note6,
  Wan.PRB.2002}.
Moreover, it is consistent with the
experimental observation that the transport ``easy axis'' is along
$B_{||}$ (i.e., $R_{xx} < R_{yy}$), as intuitively the excited
quasi-particles (at finite temperatures) should have a higher hopping
rate in the direction of charge distribution elongation.

Data taken on the 50-nm-wide QW (Fig. 4) corroborate Figs. 1-3 data
and our above conclusions, and reveal new information. In Fig. 4(a) we
show $R_{yy}$ traces at density $p = 0.95\times 10^{11}$ cm$^{-2}$ at
different angles. Qualitatively similar to the data of Figs. 1 and 2,
the traces at $\theta =0^{\circ}$ and $55^{\circ}$ appear normal and
exhibit a very strong $\nu=1$ integer QHS and numerous FQHSs at
standard fillings as well as at $\nu=1/2$. The $\theta \simeq 37^{\circ}$
trace, however, shows an IP near $\nu=1/2$. The same trace also
indicates a weak minimum near $\nu=1$ and other features, e.g. FQHSs
at $\nu=19/15$ and 29/35, indicating that the two lowest LLs are near a
coincidence. Traces taken at $\theta \simeq 35^{\circ}$ and slightly higher
density, presented in Fig. 4(b),
reveal that $R_{yy} >> R_{xx}$ and that both $R_{xx}$ and $R_{yy}$
show insulating behavior near $\nu=1/2$, similar to the
40-nm-wide QW data of Fig. 3. 

The smaller density in the 50-nm QW sample allows us to make
measurements at higher tilt angles. As seen in the top trace of
Fig. 4(a), taken at $\theta \simeq 65^{\circ}$, an IP reappears at high
$B_\perp$, past $\nu=3/5$. We believe this IP signals the onset of the
2DHS splitting into a bilayer system, similar to what is seen in 2DESs
confined to wide QWs at very large tilt angles \cite{Note7}. 

The results we report here attest to the extremely rich physics of
2DHSs confined to wide GaAs QWs. In these systems one can cause a
crossing of the lowest two LLs by either changing the density
\cite{Liu.PRB.2014} or titling the sample in magnetic field
\cite{Graninger.PRL.2011, Liu.PRB.2014,
  Liu.PRB.2015B}. Depending on the sample parameters, the
crossing can destroy the ordinary QHSs, both at integer and fractional
fillings, and bring to life unusual phases such as a FQHS at $\nu=1/2$
\cite{Liu.PRB.2014} or, as we have shown here, an anisotropic IP
signaling a two-component, striped Wigner crystal.

\begin{acknowledgments}
  We acknowledge support by the DOE BES (DE-FG02-00-ER45841) grant for
  measurements, and the NSF (Grants DMR-1305691 and MRSEC
  DMR-1420541), the Gordon and Betty Moore Foundation (Grant
  GBMF4420), and Keck Foundation for sample fabrication and
  characterization. We thank R.N. Bhatt, E. Fradkin, J.K. Jain, and S.A. Kivelson for
  illuminating discussions, and R. Winkler for providing the charge
  distribution and potential calculations shown in Figs. 1(b) and
  4(a). A portion of this work was performed at the NHMFL, which is
  supported by the NSF Cooperative Agreement No. DMR-1157490, the
  State of Florida, and the DOE. We thank S. Hannahs, G. E. Jones,
  T. P. Murphy, E. Palm, A. Suslov, and J. H. Park for technical
  assistance.
\end{acknowledgments}

\bibliographystyle{apsrev4-1}
\bibliography{../bib_full}

\end{document}